\documentclass[12pt, a4paper]{article}
\pdfoutput=1

\usepackage{amsmath}
\usepackage{amsfonts}
\usepackage{amssymb}
\usepackage{graphicx, rotating}
\usepackage{epstopdf}
\usepackage{epsfig}
\usepackage{latexsym}
\usepackage{graphicx}
\usepackage{color}
\usepackage{amsmath,amssymb}
\usepackage{cite}
\usepackage{slashed}
\usepackage{hyperref}
\usepackage{datetime}

\numberwithin{equation}{section}

\hypersetup{colorlinks, citecolor=bluscuro, linkcolor=bluscuro, urlcolor=bluscuro}
\definecolor{rossos}{rgb}{0.8,0.2,0.3}
\definecolor{bluscuro}{rgb}{0.15, 0.2, 0.9}
\definecolor{verdes}{rgb}{0.1, 0.5, 0.1}

\makeatother   % Cancel the effect of \makeatletter

\newcommand{\TeV}{{\rm \,TeV}}

 \def\be   {\begin{equation}}   \def\ee   {\end{equation}}
 \def\ba   {\begin{array}}      \def\ea   {\end{array}}
 \def\bea  {\begin{eqnarray}}   \def\eea  {\end{eqnarray}}
 \def\bean {\begin{eqnarray*}}  \def\eean {\end{eqnarray*}}

\begin{document}

\begin{flushright}
%\textit{ \dmyyyydate\today\quad\currenttime}
 \hfill
 {\footnotesize CERN-PH-TH/2011-178  \\ SACLAY-T11/167}
\end{flushright}

\vspace{0.5cm}
\begin{center}

{\Large \bf {Initial State Radiation 
in Majorana \\[0.2cm]  Dark Matter Annihilations}
}
\\[1.2cm]
{\large Paolo Ciafaloni $^{\rm a}$, Marco Cirelli $^{\rm b,c}$, Denis Comelli $^{\rm d}$,} \\[0.2cm]
{\large Andrea De Simone $^{\rm e}$, Antonio Riotto $^{\rm b,f}$, Alfredo Urbano $^{\rm a,g}$}
\\[1cm]

{\small \textit{$^{\rm a}$ Dipartimento di Fisica, Universit\`a di Lecce and INFN - Sezione di
Lecce, \\Via per Arnesano, I-73100 Lecce, Italy }}

{\small \textit{$^{\rm b}$  CERN, PH-TH Division, CH-1211,
Gen\`eve 23,  Switzerland}}

{\small \textit{$^{\rm c}$ Institut de Physique Th\'eorique, CNRS URA 2306\\ and CEA/Saclay, F-91191
 Gif-sur-Yvette, France }}

{\small \textit{$^{\rm d}$  INFN - Sezione di Ferrara, Via Saragat 3, I-44100 Ferrara, Italy}}

{\small \textit{$^{\rm e}$ Institut de Th\'eorie des Ph\'enom\`enes Physiques,\\
 \'Ecole Polytechnique F\'ed\'erale de Lausanne, CH-1015 Lausanne,
Switzerland}}

{\small \textit{$^{\rm f}$  INFN, Sezione di Padova, Via Marzolo 8, I-35131, Padova, Italy}}

{\small \textit{$^{\rm g}$  IFAE, Universitat Aut\`onoma de Barcelona, 08193 Bellaterra, Barcelona, Spain}}

\vspace{.2cm}

\end{center}

\vspace{.8cm}

\begin{center}
\textbf{Abstract}
\end{center}

\begin{quote}
The cross section for a Majorana Dark Matter particle annihilating into light fermions is helicity suppressed.
We show that, if the Dark Matter is the neutral Majorana component of a multiplet which is charged under the
 electroweak interactions of the Standard Model, the emission of gauge bosons from the initial  state 
lifts the suppression and allows an $s$-wave annihilation.
The resulting  energy spectra of stable Standard Model particles are importantly affected.
This has an impact  on  indirect searches for Dark Matter.
\end{quote}

\def\thefootnote{\arabic{footnote}}
\setcounter{footnote}{0}
\pagestyle{empty}

\newpage
\pagestyle{plain}
\setcounter{page}{1}

%%%%%%%%%%%%%%%%%%%%%%%%%%%%%%%%%%%%%
\section{Introduction and setup}
\noindent
%%%%%%%%%%%%%%%%%%%%%%%%%%%%%%%%%%%%%

\noindent
The fluxes of stable Standard Model particles that originate from the annihilation 
(or decay) of Dark Matter (DM) in the galactic halo 
are the primary observable for DM indirect searches.
The radiation of ElectroWeak (EW) gauge bosons from the final state of the annihilation
process  turns out to have a great influence on
the energy spectra of stable particles and hence on the predictions for fluxes to be measued at Earth 
\cite{paper1, paper2}  (see also Refs.~\cite{w, ibarra} for related analyses).
In particular, there are three situations where the effect of including the EW corrections is especially important: 
\begin{enumerate}
\item  when the  low-energy regions of the spectra, which are largely populated by the decay products of the emitted gauge bosons, are the ones contributing the most to the 
observed fluxes of stable particles;
\item  when some particle species are absent if EW corrections are not included, e.g.~antiprotons from 
$W/Z$ decays in an otherwise purely leptonic channel;
\item 
when the $2\to3$ annihilation cross section, with soft gauge boson emission, is comparable or even dominant with respect to the $2\to 2$ cross section.
\end{enumerate}
A possible realization of the latter condition has been studied in Ref.~\cite{paper2}, where we considered the DM as a gauge-singlet 
Majorana particle $\chi$ 
of mass $M_\chi$ annihilating into light fermions $f$ of mass $m_f\ll M_\chi$; 
it is well known
that in this case the $\chi\chi \to f\bar f$ cross section  is  suppressed. 
In fact,  one can perform the usual expansion of the cross section 
\begin{equation}
\label{expansion}
v \sigma  =a+b\,v^{2}+\mathcal{O}(v^4)\ ,
\end{equation}
where $v\sim 10^{-3}$ is the relative velocity (in units of $c$) of the DM particles in our galaxy today.
The first term, which corresponds to the annihilation of particles in a state with  $L=0$ orbital momentum
($s$-wave), is constrained by helicity arguments to be proportional to  $(m_f/M_\chi)^2$,
and hence very small for light final state fermions.
The second term, which corresponds to the annihilation in the $L=1$ state ($p$-wave), suffers from the $v^2$ suppression.
For a DM particle singlet under the SM gauge group,  the radiation of EW gauge bosons from the final state and
from the internal propagator of the annihilation process 
eludes the suppressions
and opens up a potentially sizeable $s$-wave contribution to the cross section
(see Ref.~\cite{bergstrom} for the case of photon radiation
and Ref.~\cite{drees} for gluon radiation).

In this paper we point out that there is another situation realizing the condition 3 above, 
where we expect therefore the EW corrections to have a great impact. 
Having in mind the Weakly Interacting Massive Particle candidates for DM, it is natural
not to restrict oneself to $\chi$ being a gauge singlet, 
and to consider the possibility  that the DM is part of a multiplet  charged under the EW interactions.
In this case, the DM annihilates predominantly in $s$-wave into $W^+W^-$, if kinematically allowed.
However, now even the initial state of the channel $\chi\chi\to f\bar f$  can radiate a gauge boson; 
we shall show that this process also lifts the helicity suppression and contributes to the  $s$-wave cross section,
becoming competitive with the di-boson channel.

For definiteness, we assume that the DM particle is the  electrically-neutral Majorana component of
a $SU(2)_L$ triplet $\chi^a$, with hypercharge $Y=0$ (a wino-like particle)
\footnote{
Other representations of the EW gauge group can be considered, see e.g.~Ref.~\cite{mdm}.
For instance, a simple possibility consists of two non-degenerate $SU(2)_L$ doublets with opposite hypercharge 
(higgsino-like particles).
Direct detection constraints are avoided  \cite{mdm, essig} and also an interesting LHC phenomenology 
can arise in the quasi-degenerate limit \cite{pddm}. 
}.
The coupling with the $Z$-boson is absent and this DM candidate is compatible with the
direct detection limits.
We neglect the mass splitting of the components of the multiplet, which is generated by loop effects  \cite{splitting,mdm} 
and tends to make the charged components slightly heavier than the neutral one. 
The size of this splitting is typically of the order of 100 MeV for a TeV-scale DM mass.

In order to catch the relevance of initial state emission in a model-independent way, we work in an effective field theory setup and
we restrict to consider the interactions of the triplet $\chi^a$  with the SM left-handed doublet $L=(f_{1},f_{2})^T$; the most general  dimension-6 operators are
 \be
 \label{Leff}
{\cal L}_{\rm eff}=
\frac{C_{\rm D}}{\Lambda^2} \delta_{ab}
   \left(
\bar{L}\,\gamma_\mu  P_L\, L\right)
\left(
 \bar{\chi}^a\gamma^\mu\gamma_5 \chi^b
 \right)
+ i \; \frac{C_{\rm ND}}{\Lambda^2}\epsilon_{abc}
 \left(
\bar{L}\,\gamma_\mu  P_L\sigma^c\,L\right)
\left(
 \bar{\chi}^a\gamma^\mu \chi^b
 \right)\,,
\ee
where $C_{\rm D , ND}$ are real coefficients for diagonal and non-diagonal interactions in isospin 
space, $P_{R, L}=\frac{1}{2}(1\pm \gamma_5)$ and $\sigma^i$ are the Pauli matrices.
The assumed Majorana nature of the DM forbids some operators that would lead to an $s$-wave two-body annihilation.
The
 initial state radiation lifts the helicity suppression already
 at the level of dimension-6 operators,
unlike what happens for the final state radiation, where higher-dimensional operators are needed 
(see Ref.~\cite{paper2} for more details).
A more general effective field theory analysis will be presented in Ref.~\cite{futuro}.
The effective operators in Eq.~(\ref{Leff}) can be generated  for instance as the low-energy limit of a simple toy model
\cite{paper2, ma}
where the DM interacts with
the SM left-handed fermions through the exchange of a heavy scalar doublet $\Phi$
\be
{\cal L}_{\rm int}=-y_L\;
\bar{L}\;\sigma_a\;\chi^a \;\Phi+{\rm h.c.}\,;
\ee
integrating out the scalar sector $M_{\Phi}\gg M_\chi,\,m_{W}$, one obtains the operators
in Eq.~(\ref{Leff}) with 
$C_{\rm ND}/\Lambda^2=-C_{\rm D}/\Lambda^2=|y_L|^2 /(4 M_\Phi^2)$.

In the next section we shall discuss the velocity dependence of the amplitudes describing the DM annihilation  into light fermions, considering both the Final State Radiation (FSR) and the Initial State Radiation (ISR) contributions. 
Subsequently, in section \ref{sec:results} we shall present the results for the
cross sections and the energy spectra of final particles. Our main results are summarized in 
Section \ref{sec:conclusions}, together with prospects for further research.

%%%%%%%%%%%%%%%%%%%%%%%%%%%%%%%%%%%%%
\section{Velocity dependence of the amplitude}
\label{sec:annihilation}
%%%%%%%%%%%%%%%%%%%%%%%%%%%%%%%%%%%%%

%%%%%%%%%%%%%%%%%%%
\subsection{Two-body annihilation}
\label{subsec:2body}
%%%%%%%%%%%%%%%%%%%

\begin{figure}[t]
\centering
 \includegraphics[scale=0.8]{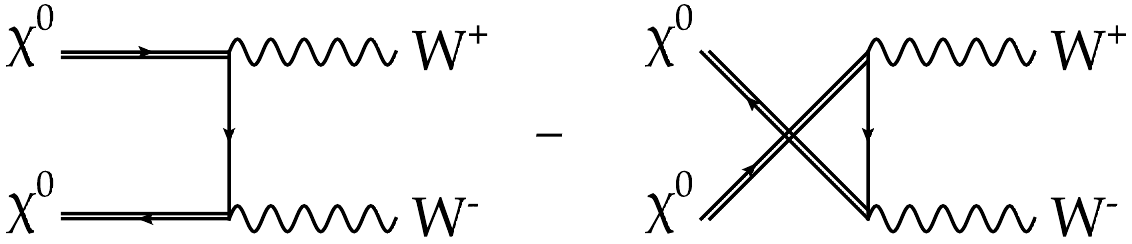}\\[0.5cm]
 \includegraphics[width=7.5cm]{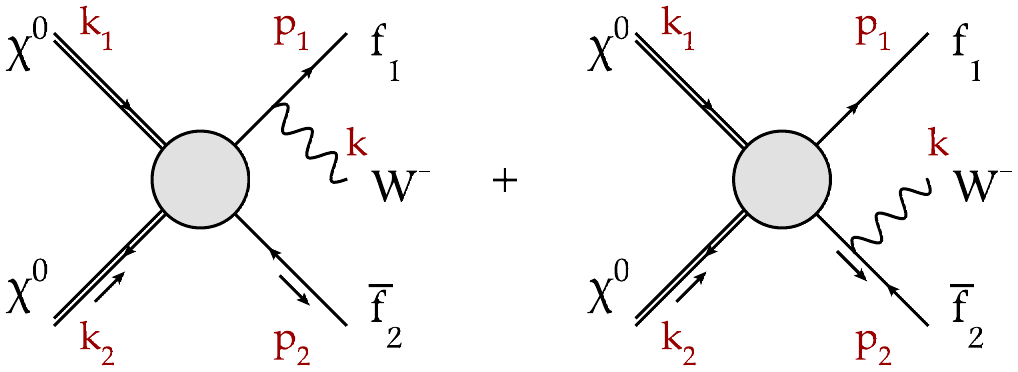} \hspace{1cm}
  \includegraphics[width=7.5cm]{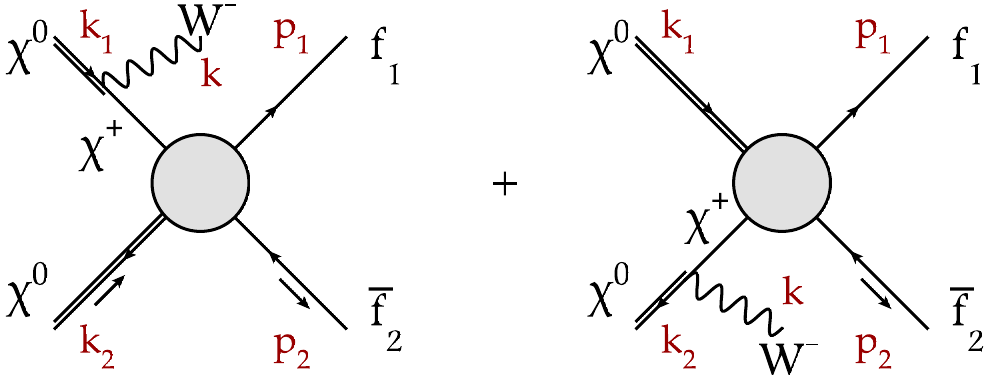}
 \caption{\small{Diagrams for the two-body annihilation into $W^+W^-$  (\emph{top panel}) and
 the three-body annihilation with
 gauge boson radiation from final (\emph{bottom left panel}) or initial (\emph{bottom right panel}) state.}}
 \label{fig:radiation}
\end{figure}

When the DM is part of a multiplet charged under the EW gauge group, and $M_\chi> m_W$, 
the most important two-body annihilation channel is $\chi^0\chi^0 \to W^+W^-$, which proceeds through $s$-wave
(see Fig.~\ref{fig:radiation}).
We shall discuss the importance of this contribution with respect to the three-body ISR channel
in  Section \ref{sec:results} and more thoroughly in Ref.~\cite{futuro}.

Let us  consider instead the annihilation of the DM Majorana fermion into a pair of left-handed  massless fermions:
\begin{equation}
\chi^0(k_1)
\chi^0(k_2)\to f_{i}(p_1)\; \bar f_{i}(p_2)\, .
\end{equation}
Only the first term of the effective operators in Eq.~(\ref{Leff}) contributes to this process
and the annihilation proceeds through the $p$-wave.
The velocity dependence is manifest at  the amplitude level. 
In fact,  the matrix element for the two-body process is 
\be
\label{amp2body}
{\cal M}_{f\bar f}\sim {1\over \Lambda^2} [\bar u_f(p_1)\,\gamma_\alpha P_L\, v_f(p_2)] \left[
\bar v_{\chi}(k_2)\,\gamma^{\alpha}\gamma_5 \,u_{\chi}(k_1)
\right]\, ,
\ee
where $k_{1,2}^{\alpha}=({M_\chi}/{\sqrt{1-(v/2)^2}},0,0,\pm
{M_\chi (v/2)}/{\sqrt{1-(v/2)^2}})$.  
The Majorana axial current in Eq.~(\ref{amp2body}) can be manipulated using the Gordon identities into
\be
\label{gordon}
\bar{v}_{\chi}(k_2)\gamma^{\alpha}\gamma_5 u_{\chi}(k_1)
=-\frac{k_1^\alpha+k_2^\alpha}{2M_\chi}
\bar{v}_{\chi}(k_2)\gamma_5 u_{\chi}(k_1)
-\frac{i}{2 M_\chi}
\bar{v}_{\chi}(k_2)\sigma^{\alpha\beta}(k_{1}-k_{2})_{\beta}\gamma_5 u_{\chi}(k_1)\,.
\ee
For small $v$,   the second term is proportional to $v$ because $(k_{1}-k_{2})^\beta \sim (0,0,0,v\,M_\chi)$; 
on the other hand, in the first term the vector $(k_1+k_2)^\alpha=(p_1+p_2)^\alpha$ saturates the final-state current in Eq.~(\ref{amp2body})
and gives rise to terms  proportional to the fermion mass, which are zero in our computation.
We thus recovered the well-known fact that  for Majorana spinors the scattering amplitude into massless 
fermions is proportional to the first power of the relative velocity of the incoming particles.

%%%%%%%%%%%%%%%%%%%
\subsection{Gauge boson emission}
\label{subsec:3body}
%%%%%%%%%%%%%%%%%%%

The gauge interactions of the fermion triplet are described by the vector operator
$\;\epsilon_{abc}\bar{\chi}^a\slashed{W}^b \chi^c$,
from which it is evident that the neutral component of $\chi$ does not interact with $W^3$.
Thus, $W^3$ is emitted only from the final states, while $W^{\pm}$ can be emitted from 
either initial or final states.
Let us study a $W$-bremsstrahlung process from either initial or final states,
see diagrams in Fig.~\ref{fig:radiation}.
We consider for definiteness the three-body annihilation
\begin{equation}
\chi^0(k_1)
\chi^0(k_2)\to f_{1}(p_1) \; \bar f_{2}(p_2) \; W^-(k).
\end{equation}
For the process with FSR, the interaction vertex of two $\chi^0$'s and two SM fermions  is described by the isospin-diagonal operator in Eq.~(\ref{Leff}); instead, since ISR changes the isospin state of one of the initial legs, the interaction operator is the non-diagonal one.

Considering the radiation from the final state, the amplitude is the product of the Majorana axial-vector current
$\bar v_{\chi}\,\gamma_\mu\gamma_5\,u_\chi$
and the fermionic  current containing the gauge boson emission
\be
\label{ampFSR}
{\cal M}_{\rm FSR}\sim  {g\over \Lambda^2} \left[\bar v_{\chi}\,\gamma_\mu\gamma_5\,u_\chi\right]
\left[
\bar{u}_{f}\left(
\frac{\slashed{\epsilon}^* (\slashed{p}_1+\slashed{k})\gamma^{\mu}}{2p_1\cdot k+m_W^2}
-\frac{ \gamma^{\mu}(\slashed{p}_2+\slashed{k}) \slashed{\epsilon}^* }{2p_2\cdot k+m_W^2}\right)
P_L v_{f}
\right]\,,
\ee
where $g$ is the $SU(2)_L$ gauge coupling.
Using the Gordon identities the Majorana current  can be simplified as in Eq.~(\ref{gordon}), where the 
second term is proportional to $v$, while  the first term now contains the 4-vector  $(p_1+p_2+k)_{\mu}$;
saturating with the fermionic current in (\ref{ampFSR}) gives terms proportional to $m_f$, which are zero in our case. 
Thus, at the lowest level in the expansion in $M_\chi^2/\Lambda^2$, which amounts to restricting to 
dimension-6 operators, FSR is not able to  remove the helicity suppression.
However, at  higher orders  (e.g. ${\cal O}(M_\chi^4/\Lambda^4)$ in the amplitude) 
 $v$-independent terms can arise, for which
 the inclusion of  Virtual Internal Bremsstrahlung (VIB)  diagrams
\cite{bergstrom} is crucial. This point is extensively discussed in Ref.~\cite{paper2}.

%\bea\nonumber
%\bar{v}_{\chi}(k_2)\gamma_{\mu}\gamma_5 u_{\chi}(k_1)
%\stackrel{v\to 0}{\longrightarrow}
%\frac{(p_1+p_2+k)_{\mu} }{2M_\chi}
%\bar{v}_{\chi}\gamma_5\; u_{\chi}+{\cal O}(v)%\quad
%%{\rm with}\quad k^{\alpha}=\lim_{v\to 0}k_{1,\,2}^{\alpha}\, .
%\eea
%and now the vector $(p_1+p_2+k)_{\mu}$ saturating the fermionic currents (\ref{ampFSR}) gives zero  terms for vanishing $m_f$
%and the FSR amplitude vanishes in the limit $v\to 0$.

On the other hand, the amplitude describing the radiation from the initial state
consists of the product of the fermionic current  $\bar u_{f}\,\gamma_\mu P_L\,v_f$
and the Majorana current with the gauge boson emission
\be
{\cal M}_{\rm ISR}\sim 
g {1\over \Lambda^2} \left[\bar u_{f}\,\gamma_\mu P_L\,v_f\right]
\left[\bar v_{\chi}\left(
\frac{\slashed{\epsilon}^*
(\slashed{k}-\slashed{k}_2+M_\chi)\gamma^\mu}
{m_W^2-2k\cdot k_2}+
\frac{\gamma^\mu \,
(\slashed{k}_1-\slashed{k}+M_\chi)\slashed{\epsilon}^* }{m_W^2-2k\cdot k_1} \right)
u_{\chi}
\right]\,.
\ee
The effect of ISR is to alter the axial-vector structure of the initial state current, 
thus preventing the amplitude from vanishing in the $v\to 0$ limit
\be
{\cal M}_{\rm ISR}^{(v=0)}\sim 
{g\over \Lambda^2} \left[\bar u_{f}\,\gamma_\mu P_L\,v_f\right]
\left[
\bar v_{\chi}
\frac{\{\slashed{\epsilon}^* \slashed{k},\gamma^{\mu}\}}
{m^2-2k^0 M_\chi}
u_{\chi}
 \right]\,.
\label{ampISR}
\ee
From the amplitudes we studied in this section one can deduce the  behaviours of the cross
sections for the  two-body $f\bar f$ channel and the three-body channels with ISR, FSR and their interference;
they can be schematically summarized as
\bea
v\sigma_{f\bar f}(\chi\chi\to f_i \bar f_i)&\sim&
{1\over M_\chi^2}{\cal O}\left( \frac{M_\chi^4}{\Lambda^4}\right)\,{\cal O}(v^2)\,, \\
v\sigma_{\rm FSR, \;ISR/FSR}(\chi^0\chi^0\to f_1 \bar f_2 W^-)&\sim&
{g^2\over M_\chi^2}{\cal O}\left( \frac{M_\chi^4}{\Lambda^4}\right) {\cal O}(v^2)\,, 
\label{MFSR}\\
v\sigma_{\rm ISR}(\chi^0\chi^0\to f_1 \bar f_2 W^-) &\sim&
{g^2\over M_\chi^2}{\cal O}\left( \frac{M_\chi^4}{\Lambda^4}\right)\, {\cal O}(v^0)\,,
\eea
and the precise expressions  will be discussed in the next section.
The importance of ISR is then clear: at the level of dimension-6 operators 
ISR already opens  the $s$-wave annihilation while  FSR is still in $p$-wave.

%\begin{small}
%As already shown in \cite{paper2} - in fact -
%an order-of-magnitude estimate for the FSR cross section can be obtained straightforwardly
%%
%\bea
%v\sigma(\chi\chi\to f_i \bar f_i)&\sim& {1\over M_\chi^2}\,
% {\cal O}\left(\frac{M_\chi^4}{\Lambda^4}\right)\,{\cal O}(v^2)\,, \nn \\
%v\sigma^{\rm FSR}(\chi^0\chi^0\to f_1 \bar f_2 W^-)&\sim& {\alpha_W\over M_\chi^2}\,
%\left[
% {\cal O}\left(\frac{M_\chi^4}{\Lambda^4}\right) {\cal O}(v^2)+
% {\cal O}\left(\frac{M_\chi^6}{\Lambda^6}\right) {\cal O}(v^2)+
% {\cal O}\left(\frac{M_\chi^8}{\Lambda^8}\right) {\cal O}(v^0)
% \right]\,, \nn\\
%%\frac{1}{\Lambda^2}\left[
%% {\cal O}(v^2)\;\frac{M_\chi^2}{\Lambda^2}+
%% {\cal O}(v^2)\;\frac{M_\chi^4}{\Lambda^4}+ {\cal O}(1)\;\frac{M_\chi^6}{\Lambda^6}
%%\right] 
%v\sigma^{\rm ISR}(\chi^0\chi^0\to f_1 \bar f_2 W^-) &\sim&{\alpha_W\over M_\chi^2}\,
% {\cal O}\left(\frac{M_\chi^4}{\Lambda^4}\right)\, {\cal O}(v^0)\,,
%\eea
%%
%where $\alpha_W=g^2/(4\pi)$.
%\end{small}

%%%%%%%%%%%%%%%%%%%%%%%%%%%%%%%%%%%%%
\section{Results}
\label{sec:results}
%%%%%%%%%%%%%%%%%%%%%%%%%%%%%%%%%%%%%

%Let us now turn to the results for the cross section and for the energy distributions  of final particles.

%%%%%%%%%%%%%%%%%%%
\subsection{Cross sections}
%%%%%%%%%%%%%%%%%%%

The cross sections for the various processes are computed from the amplitudes in Eqs.~(\ref{amp2body}), (\ref{ampFSR}), 
(\ref{ampISR}), with the appropriate coefficients  as dictated by the interaction lagrangian (\ref{Leff}).
For the two-body annihilation into massless fermions, the cross section reads 
\be
v\sigma_{f\bar f}=\frac{C_{\rm D}^2}{12\pi}\frac{M_\chi^2 }{\Lambda^4}\, v^2 \, .
\ee
For the three-body processes, it is convenient to define $s\equiv (k_1+k_2)^2$ and   $z\equiv m_W/\sqrt{s}$. 
We are working in a situation where the DM mass is larger
than the EW scale; so we report results as an expansion for small $z$.
Then, the cross sections for the processes with FSR and with ISR/FSR interference are
\bea
v\sigma_{\rm FSR}&=&
\frac{g^2C_{\rm D}^2}{144\pi^3}{M_{\chi}^2\over \Lambda^4}\left[
15-\pi^2+6 \ln z\left(2\ln z+3\right) +{\cal O}(z^2) \right] v^2\,. 
\label{eq:FSRtotale}\\
v\sigma_{\rm ISR/FSR}&=&
-\frac{g^2C_{\rm D}C_{\rm ND}}{16\pi^3}{M_{\chi}^2\over \Lambda^4}
\left[9+4\ln z + {\cal O}(z)\right] v^2\,,
\label{eq:ISRFSRtot}
\eea
which are both in $p$-wave. 
As recalled already in the previous section,  an $s$-wave term in the cross section also originates from 
FSR (together with VIB)  but only at a higher order, $O(M_{\chi}^8/\Lambda^8)$ in the cross section. 
Nonetheless, this can have a great impact on the energy spectra of final particles \cite{paper2},
but we shall not consider it here.
Instead the ISR  opens up a large $s$-wave contribution to the total cross section
\begin{equation}\label{eq:totalISR}
v\sigma_{\rm ISR}=\frac{g^2 C_{\rm ND}^2 }{9 \pi^3}{M_\chi^2\over  \Lambda^4}\,,
\end{equation}
which is mediated by the non-diagonal operator in Eq.~(\ref{Leff})
\footnote{
The isospin-diagonal operator leads to a  spin-dependent elastic DM-nucleon cross section
proportional to  $C_{\rm D}^2$.
Experimental bounds from direct detection would therefore  constrain $C_{\rm D}$,
but would not  preclude  {\it a priori}  the possibility to have a large $\sigma_{\rm ISR}$.
}.
%
%Let us set the energy of the $W$ as $k^0=E_W\equiv (1-y)\sqrt{s}/2$.
%\begin{eqnarray}
% k^0 &=& (1-x_2)\sqrt{s_1}/2\, , \label{eq:k}
% p_1^0 &=& x_1\sqrt{s_1}/2\, ,\label{eq:p1} \\
 %p_2^0 &=& (1-x_1+x_2)\sqrt{s_1}/2\, ,\label{eq:p2}
%The   variable  $y$ is subject to the kinematical constraint $-z^2 \leq  y \leq 1-2z$.

As already anticipated, there is also an important two-body channel 
$\chi^0\chi^0\to W^+ W^-$, whose cross
section is not $v$-suppressed
\be
v\sigma_{\rm WW} = \frac{g^4}{8\pi M_{\chi}^2}+{\cal O}(z^4)\,.
\ee
and can be comparable in size to $\sigma_{\rm ISR}$. 
The annihilation channel with ISR can even become the dominant one if 
for instance the $W^+W^-$ final state is kinematically forbidden, or if the coefficient $C_{\rm ND}$ is large.
Notice also that the additional $g^2$ factor in  $\sigma_{\rm WW}$ tends to reduce it with respect
to $\sigma_{\rm ISR}$, if $C_{\rm ND}\sim 1$;
furthermore,
the number of colours or families of the final state fermions, which we do not consider here,  
also enhances $\sigma_{\rm ISR}$ with respect to $\sigma_{\rm WW}$.

%%%%%%%%%%%%%%%%%%%
\subsection{Energy spectra}
%%%%%%%%%%%%%%%%%%%

Let us now turn to the phenomenological implications of ISR,
namely the effect of the $W$-bremsstrahlung on the energy
spectra of stable particles  resulting from the hadronization and decay
of the annihilation products.

In this paper we are interested in extracting the features distinguishing ISR from other
kind of processes, like the two-body $W^+W^-$ channel and the three-body FSR,  
although their cross section can be very different.
To make the comparisons more immediate,
we define the energy spectrum of the channel $i$ as
\begin{equation}\label{eq:norm}
{dN_i\over dE}\equiv \frac{1}{\sigma_i}
{d\sigma_i\over dE}\,, 
\end{equation}
for $i=$ FSR, ISR, ISR/FSR, WW,  so that each contribution is normalized to 1.
A more detailed analysis will be presented in Ref.~\cite{futuro}, where 
we shall work out an explicit model and  weigh the various channels appropriately.

\begin{figure}[t]
\centering
 \includegraphics[scale=0.7]{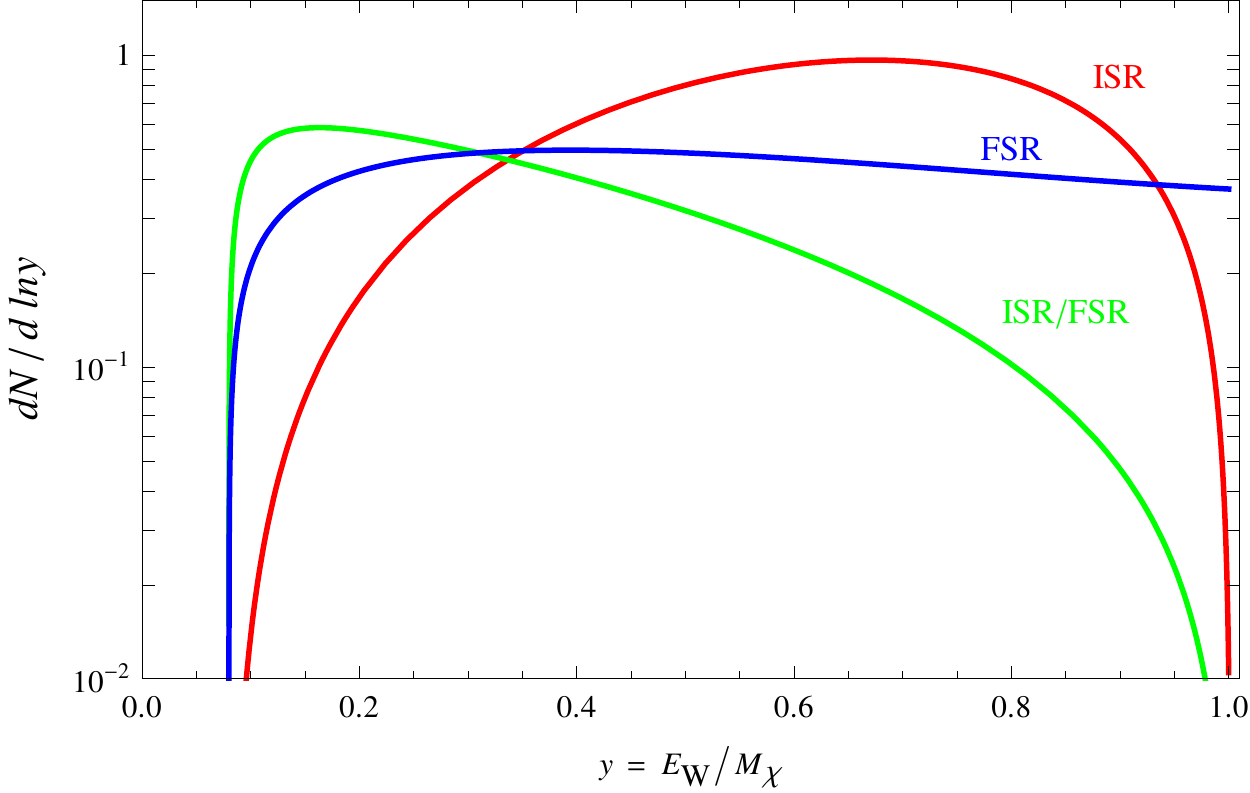}\\
 \caption{ \small{Energy spectrum of the emitted $W$ for the three contributions FSR, ISR, ISR/FSR.
  The parameters settings are $M_\chi=1 \TeV$,  $v=10^{-3}$. Each distribution is normalized to 1 
  according to Eq.~(\ref{eq:norm}).
  }}
 \label{fig:Wspectra}
\end{figure}

Because  the energy of the emitted $W$ boson is entirely distributed
among the final particles, it is instructive to analyse its energy spectrum separating the 
contributions from ISR, FSR and their interference, see Fig.~\ref{fig:Wspectra}.
Notice that the $W$ emission
from an ultra-relativistic final state particle (FSR) has a characteristic
soft/collinear behaviour $dN/dy\sim 1/y$, where $y=E_W/M_\chi$
(see Ref.~\cite{paper1} for further details); 
on the other hand, 
the $W$ emission from a non-relativistic initial state particle (ISR)
shows a somehow peculiar energy spectrum, which turns out to be well
approximated by the symmetric distribution $dN/dy\sim y(1-y)$. 
Due to the non-relativistic nature of the emitting DM particle
and the failure of the factorization property for the three-body cross
section, this contribution cannot be caught by the usual
soft/collinear approximation technique \cite{paper1}.

The evolution of the particle species from the primary annihilation products to the final 
stable particles of the SM needs numerical tools. We have carried out this work using 
our own Monte Carlo
code, for generating $2\to2$ and $2\to 3$ annihilation events, then interfaced to 
 {\sc Pythia} 8.145 \cite{pythia} for simulating the subsequent showering,
hadronization and decay (see Ref.~\cite{paper2}  for more details).

\begin{figure}[t]
\centering
\includegraphics[width=7.5cm]{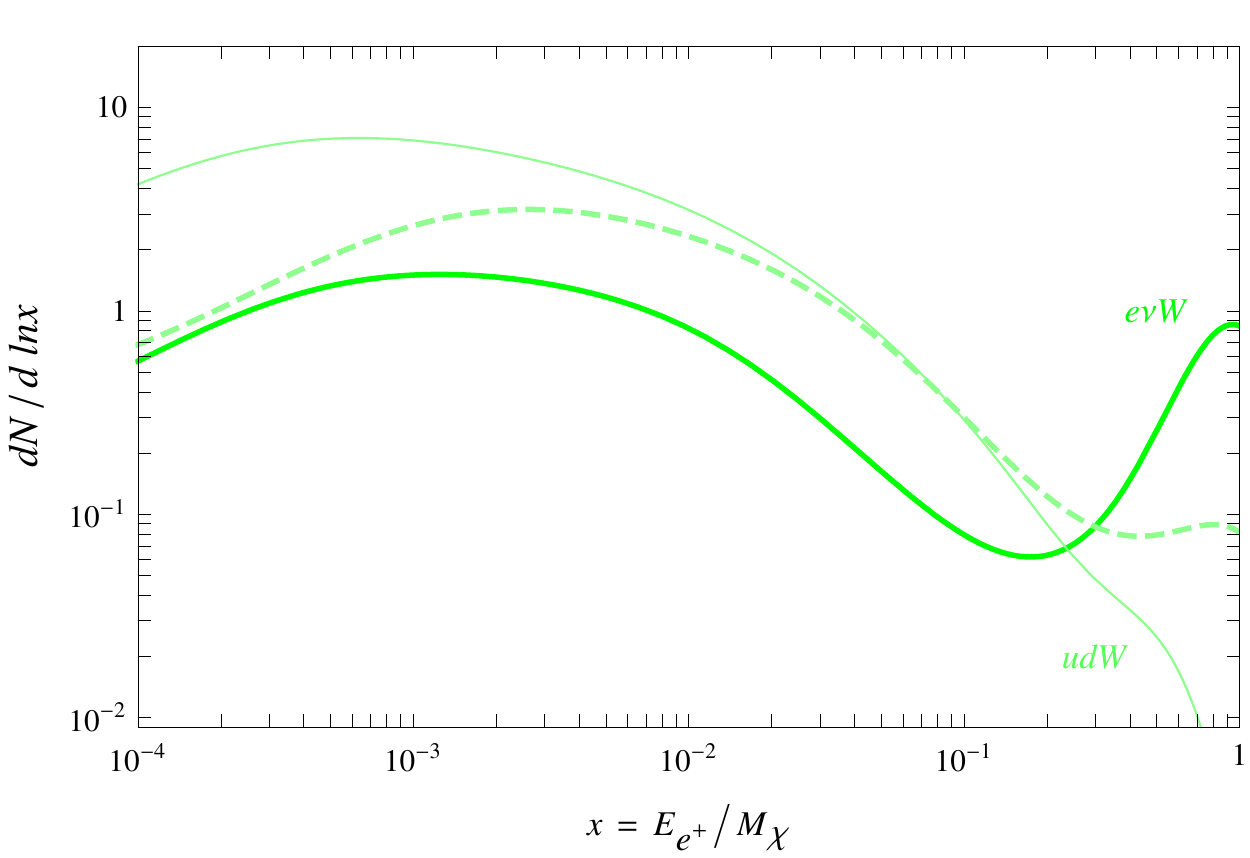}
 \hspace{1cm}
\includegraphics[width=7.5cm]{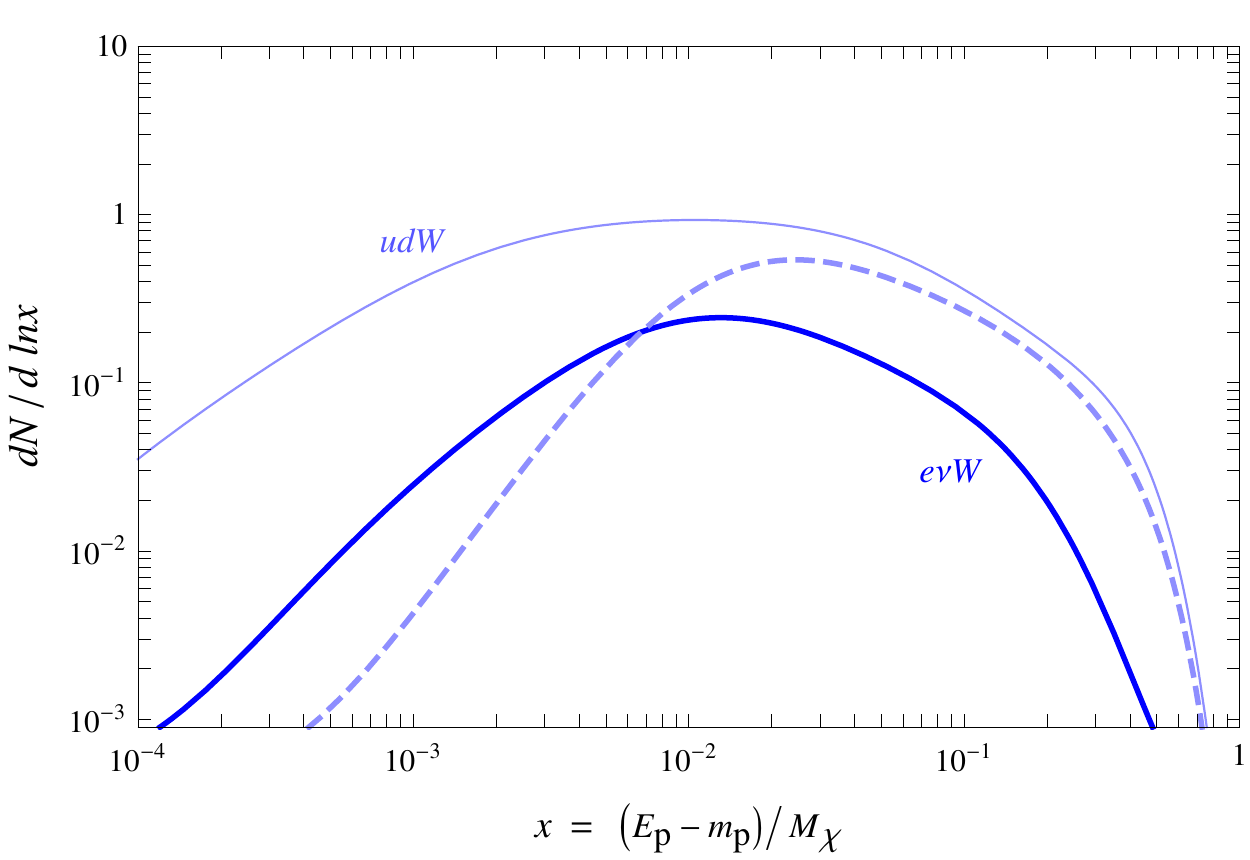}
 \caption{\small{Energy spectra of final positrons (\emph{left panel}) and antiprotons (\emph{right panel}),
for different annihilation channels defined in the text: $W^+W^-$ (dashed lines), $e\nu W$ (thick solid lines)
and $udW$ (thin solid lines).
The parameters are set as $M_\chi=1 \TeV$ and  $v=10^{-3}$, and the normalization is chosen 
according to Eq.~(\ref{eq:norm}).
 }}
 \label{fig:confronto}
\end{figure}

In  Fig.~\ref{fig:confronto},  we compare the energy spectra of final positrons and antiprotons
originating from ISR and from WW. We consider two different channels for the ISR case:
the lepton channel ``$e\nu W$'': $\chi^0\chi^0\to e_L^+ \nu_{e\,L} W^-$, $e_L^- \bar\nu_{e\,L} W^+$ ;
and the
quark channel ``$udW$'': $\chi^0\chi^0\to u_L \bar d_L W^-$, $\bar u_L d_L W^+ $ .
Interesting features  can be extracted from this comparison.
For the  $e\nu W$  channel
the very hard positrons can be much more abundant for ISR than for WW,
due the contribution of the primary positrons. 
For the $udW$ channel,
the antiprotons are copiously generated by the $W$ emitted in ISR, especially at low energies
because the gauge boson is soft, and by the hadronization of the primary quarks,
and they can easily overcome the antiprotons produced in the WW channel.

The fluxes of positrons and antiprotons received at Earth can be computed by integrating the
 energy spectra at the interaction point depicted in Fig.~\ref{fig:confronto} over the diffused source
 constituted by the DM distribution in the galactic halo,  and then propagating them through the halo itself.
We do not enter here into any of the details of this process (see e.g.~Ref.~\cite{pppc} and references therein) but we limit ourselves to a few qualitative considerations.
There are irreducible uncertainties of astrophysical nature that originate from: i) the unknown distribution of DM in the halo; ii) the unknown values of the propagation parameters.
For the positron fluxes at high energy, however, the propagation does not sensibly modify the shape and the normalization of the fluxes, for any reasonable choice of the uncertain variables. This is just because high energy positrons have anyway originated very close to the location of the Earth and therefore have not been affected much during their travel. In turn, this implies that the spectral features at high energy introduced by ISR, apparent in Fig.~\ref{fig:confronto}, are preserved in the final positron fluxes, at least in the case of large DM mass. On the other hand, the fluxes of antiprotons are affected by large astrophysical uncertainties both in normalization and in shape all across the energy range, making it more difficult to disentangle the different spectral shapes.

%%%%%%%%%%%%%%%%%%%%%%%%%%%%%%%%%%%%%
\section{Summary and outlook}
\label{sec:conclusions}
%%%%%%%%%%%%%%%%%%%%%%%%%%%%%%%%%%%%%

The inclusion of EW corrections is an essential ingredient to be taken into account
for indirect DM searches. In this paper 
we have assumed that the  DM  is the neutral Majorana component of a multiplet charged
under the EW interactions
 and considered the effect of  gauge boson radiation
from the initial state of the DM annihilation process.
We have restricted ourselves to the case where the multiplet containing the DM particle
is a $SU(2)_L$ triplet,
but it is straightforward to work out  the cases of
multiplets transforming under different representations  of the  EW gauge group.

The natural annihilation channel for such a candidate is of course through $s$-wave into $W^+ W^-$, if  kinematically allowed,
while the annihilation into light SM fermions is helicity suppressed and proceeds through $p$-wave.
However, we found that the $W$-bremsstrahlung from the initial state  removes the suppression
and adds a  potentially sizeable contribution to the  $s$-wave cross section.
The gauge boson emission alters the energy spectra of final stable particles in a distinguishable
way and cannot be ignored for reliable predictions to be used for indirect DM searches.

In a forthcoming paper \cite{futuro} we shall expand the idea of this work in
several directions: a complete effective field theory analysis, a full calculation 
in the context of an explicit model which will also allow to weigh precisely
the different
channels, a detailed computation of fluxes of stable particles with the 
inclusion of the propagation effects.

%%%%%%%%%%%%%%%%%%%%%%%%%%%%%%%%%%%%%
\section*{Acknowledgments}
%%%%%%%%%%%%%%%%%%%%%%%%%%%%%%%%%%%%%
The work of ADS is supported by the Swiss National Science Foundation under
contract 200021-125237.
The work of AU is supported by CICYT-FEDER-FPA$2008$-$01430$.
The work of MC is supported in part by the French national research agency 
ANR under the contract ANR 2010 BLANC 041301 ``TH-EXP@TeV''
and by the ITN network UNILHC.

%%%%%%%%%%%%%%%%%%%%%%%%%%%%%%%%%%%%%
\bibliographystyle{JHEP}

%%%%%%%%%%%%%%%%%%%%%%%%%%%%%%%%%%%%%

\end{document}